\documentclass[onecolumn,longauth]{aa} 

\usepackage{graphicx}
\usepackage{txfonts}
\usepackage{ulem}
\usepackage[colorlinks=true,citecolor=blue]{hyperref}
\makeatletter
\renewcommand*\aa@pageof{, page \thepage{} of \pageref*{LastPage}}
\makeatother

\bibpunct{(}{)}{;}{a}{}{,}

\begin{document} 

\title{Searching for the near infrared counterpart of Proxima c using multi-epoch high contrast SPHERE data at VLT \thanks{Based on data collected at the European Southern Observatory, Chile (ESO Programs 095.D-0309, 096.C-0241, 096.D-0252, 097.C-0865, 198.C-D0209, 099.D-0098, 099.C-0127}}

   \author{R. Gratton\inst{1}
          \and
          A. Zurlo\inst{2,3,4}
          \and
          H. Le Coroller\inst{4}
          \and
          M. Damasso\inst{5}
          \and
          F. Del Sordo\inst{6,7}
          \and
          M. Langlois\inst{8,4}
          \and
          D. Mesa\inst{1}
          \and
          J. Milli\inst{9}
          \and
          G. Chauvin\inst{10,11}
          \and
          S. Desidera\inst{1}
          \and
          J. Hagelberg\inst{12}
          \and
          E. Lagadec\inst{13}
          \and
          A. Vigan\inst{4}
          \and
A. Boccaletti\inst{14}
\and 
M. Bonnefoy\inst{10}
\and 
W. Brandner\inst{15} 
\and 
S. Brown\inst{15}
\and 
F. Cantalloube\inst{15} 
\and
P. Delorme\inst{10}
\and 
V. D'Orazi\inst{1} 
\and 
M. Feldt\inst{15} 
\and 
R. Galicher\inst{14} 
\and
T. Henning\inst{15}
\and
M. Janson\inst{16}
\and
P. Kervella\inst{13}
\and 
A-M. Lagrange\inst{10} 
\and 
C. Lazzoni\inst{1}
\and 
R. Ligi\inst{17}
\and 
A-L. Maire\inst{15,18}
\and
F. M\'enard\inst{10}
\and 
M. Meyer\inst{19}
\and 
L.Mugnier\inst{20}
\and
A. Potier\inst{14} 
\and
E. L. Rickman\inst{12}
\and
L. Rodet\inst{10} 
\and 
C. Romero\inst{10}
\and 
T. Schmidt\inst{14,21} 
\and
E. Sissa\inst{1}
\and
A. Sozzetti\inst{5}
\and
J. Szul\'{a}gyi\inst{22}
\and
Z. Wahhaj\inst{9}
\and
J. Antichi\inst{23}
\and
T. Fusco\inst{20}
\and
E. Stadler\inst{10}
\and
M. Suarez\inst{24}
\and
F. Wildi\inst{12}
}

   \institute{INAF - Osservatorio Astronomico di Padova, Italy
              \email{raffaele.gratton@inaf.it}
              \and
    N\'ucleo de Astronom\'ia, Facultad de Ingenier\'ia y Ciencias, Universidad Diego Portales, Av. Ejercito 441, Santiago, Chile 
    \and
    Escuela de Ingenier\'ia Industrial, Facultad de Ingenier\'ia y Ciencias, Universidad Diego Portales, Av. Ejercito 441, Santiago, Chile 
              \and
    Aix Marseille Universit\'e, CNRS, LAM (Laboratoire d'Astrophysique de Marseille) UMR 7326, 13388 Marseille, France    \and
    INAF - Osservatorio Astrofisico di Torino, Italy
              \and
    Institute of Astrophysics, FORTH, GR-71110 Heraklion, Greece
    \and
    Department of Physics, University of Crete, GR-70013 Heraklion, Greece
            \and
 CRAL, UMR 5574, CNRS, Université Lyon 1, ENS, 9 avenue Charles André, 69561 Saint Genis Laval Cedex, France            
    \and
    European Southern Observatory, Alonso de Cordova 3107, Casilla 19001 Vitacura, Santiago 19, Chile 
    \and
    Univ. Grenoble Alpes, CNRS, IPAG, F-38000 Grenoble, France
    \and
    Unidad Mixta Internacional Franco-Chilena de Astronom\'ia, CNRS/INSU UMI 3386 and Departamento de Astronomía, Universidad de Chile, Casilla 36-D, Santiago, Chile
    \and
    Observatoire de Gen\`eve, Universit\'e de Gen\`eve, 51 Chemin des Mailletes, 1290 Sauverny, Switzerland
    \and
    Universit\'e C\^ote d’Azur, OCA, CNRS, Lagrange, France
    \and 
    LESIA, Observatoire de Paris, Universit\'e PSL, CNRS, Sorbonne Université, Univ. Paris Diderot, Sorbonne Paris Cit\'e, 5 place Jules Janssen, 92195 Meudon, France 
    \and
    Max Planck Institute for Astronomy, K\"onigstuhl 17, D-69117 Heidelberg, Germany,
    \and
    Department   of   Astronomy,   Stockholm   University,   Stockholm,Sweden
    \and
    INAF-Osservatorio Astrnomico di Brera
    \and
    STAR Institute, Universit\'e de L\'iege, All\'ee du Six Ao\'ut 19c, B-4000,Li\'ege, Belgium
    \and
Department of Astronomy, University of Michigan, 1085 S. Univer-sity Ave, Ann Arbor, MI 48109-1107, USA
\and
    ONERA (Office National d’Etudes et de Recherches A\'erospatiales), B.P.72, F-92322 Chatillon, France
\and
Hamburger  Sternwarte,  Gojenbergsweg  112,  D-21029  Hamburg, Germany
\and
Center for Theoretical Astrophysics and Cosmology, Institute for Computational Science, University of Z\"urich, Z\"urich, Switzerland
\and
INAF - Osservatorio Astrofisico di Arcetri, Largo E. Fermi 5, I-50125 Firenze, Italy
\and European Southern Observatory, Karl-Schwarzschild-Strasse 2, D-85748 Garching, Germany
}

   \date{Received ; accepted }

 
  \abstract
   {Proxima Centauri is the closest star to the Sun and it is known to host an earth-like planet in its habitable zone; very recently a  second candidate planet was proposed based on radial velocities. At quadrature, the expected projected separation of this new candidate is larger than 1 arcsec, making it a potentially interesting target for direct imaging.}
   {While expected to be very difficult, identification of the optical counterpart of this planet would allow detailed characterization of the closest planetary system. }
   {We searched for a counterpart in SPHERE images acquired during four years through the SHINE survey. In order to account for the expected large orbital motion of the planet, we used a method that assumes the circular orbit obtained from radial velocities and exploits the sequence of observations acquired close to quadrature in the orbit. We checked this with a more general approach that considers keplerian motion, K-stacker.}
   {We did not obtain a clear detection. The best candidate has S/N=6.1 in the combined image. A statistical test suggests that the probability that this detection is due to random fluctuation of noise is $<1$\% but this result depends on the assumption that distribution of noise is uniform over the image, a fact that is likely not true. The position of this candidate and the orientation of its orbital plane fit well with observations in the ALMA 12m array image. However the astrometric signal expected from the orbit of the candidate we detected is $3 \sigma$ away from the astrometric motion of Proxima as measured from early \textit{Gaia} data. This, together with the unexpectedly high flux associated with our direct imaging detection, means we cannot confirm that our candidate is indeed Proxima c.}
   {On the other hand, if confirmed, this would be the first observation in imaging of a planet discovered from radial velocities and the second one (after Fomalhaut b) of reflecting circumplanetary material. 
   Further confirmation observations should be done as soon as possible.}


   \maketitle
%

\section{Introduction}

Proxima Centauri (hereafter Proxima) is the closest star to the Sun ($1.3012\pm 0.0003$~pc: \citealt{Benedict1999, vanLeeuwen2007, Gaia2018}) and its planetary system is among the most likely to allow a detailed investigation of an Earth-like planet. \citet{AngladaEscude2016} discovered a close-in Earth-like planet, Proxima b, using radial velocities (RV). This planet was confirmed by \citet{DD2017} and a more accurate estimate of the mass of $m_{\rm b}~\sin{i_{\rm b}}=1.0\pm 0.1$~M$_\oplus$ was obtained by \citet{Damasso2020}. This planet is in the habitable zone, but too close to the star for direct imaging with current instrumentation (projected semi-major axis $a\sim 37$~mas). In the near future, a combination of high-resolution spectroscopy and high-contrast imaging might allow detecting its signal and ultimately studying the composition of its atmosphere \citep{Snellen2015, Lovis2017}. Through additional RVs, \citet{Damasso2020} found evidence of a second planet (Proxima c) with a minimum mass of $5.8\pm 1.9$~M$_\oplus$\ on a roughly circular orbit with period of $1900^{+96}_{-82}$\ days=$5.21^{+0.26}_{-0.22}$~yr, and semi-major axis of $1.48\pm 0.08$~au, corresponding to a maximum angular separation of $1.14\pm 0.06$~arcsec. This planet is compatible with the upper limit set from astrometry \citep{Benedict1999, Lurie2014}. Hereafter, we will use the epoch, period and semi-major axis of the circular orbit solution by \citet{Damasso2020} to fine tune our search for Proxima c.

Given its large apparent separation from the star, direct detection of Proxima c might be perhaps feasible though difficult due to the large expected contrast. In this paper we report on the search of the optical counterpart of the candidate planet Proxima~c in a series of observations acquired from the SPHERE \citep{Beuzit2019} Guaranteed Time Observations. This data set was obtained with the aim of measuring the mass of Proxima from the relativistic deflection of the apparent motion of background stars \citep{Zurlo2018}. \citet{Mesa2017} performed an analysis of the limits in the mass of possible companions from data available at the time. The mass limit obtained by \citet{Mesa2017} is an order of magnitude or more higher than the most probable mass for the candidate planet proposed by \cite{Damasso2020}. However, there are a number of additional facts that justify a reanalysis of the data: (i) \cite{Mesa2017} used only a subset of the data available now. We acquired later new data sets that were obtained in better atmospheric conditions and are therefore of higher quality. Furthermore, given the very large expected orbital motion, \cite{Mesa2017}  did not combine results from different epochs and used only the best data set for their analysis. We can now try to combine results obtained at different epochs in a deeper search. (ii) Since there was no candidate at the time, the search was completely blind. In order to reduce the False Alarm Probability (FAP), \cite{Mesa2017} had to adopt a conservative detection threshold that is the usual 5-$\sigma$ level used in direct imaging surveys. Availability of a spectroscopic orbit allows a very significant reduction of the search area, relaxing this condition substantially. (iii) \cite{Mesa2017} considered only planets shining by their internal energy; while reflection of stellar light by the planet itself is not expected to be larger than a few $10^{-9}$ for the case of Proxima c, it might be enhanced e.g. by the presence of debris circumplanetary rings (see e.g. \citealt{Arnold2004}) or dust around the planet as proposed for Fomalhaut b \citep{Kalas2008, Kennedy2011}. Thanks to the additional knowledge provided by the spectroscopic obit and considering the more specific cases described above we endeavor to reanalyse the detectability of this counterpart. As shown in the remaining part of this paper, current high contrast data still do not provide a robust detection of Proxima~c. However, since for practical reasons we have no opportunity to observe again Proxima with high contrast imagers in 2020, we present our analysis in a paper because we think it may be useful to others, e.g. in preparation and comparison with future high contrast imaging, Gaia or ALMA observations. In this paper we also describe methods that can be useful in similar analysis for this and other objects.

Given the distance of Proxima and the length of the proposed period for Proxima~c, the planet is expected to move rapidly along its orbit around the star (a few mas/day). This must be taken into account when combining results from different epochs. We considered two possible approaches. The first one uses the slower apparent motion of the candidate close to expected orbital quadrature, and is similar to that considered by \citet{Mawet2019} in their search for the planet around $\epsilon$~Eri; the first guess on the orbital properties obtained with this approach is then used as a guide for identification of the possible planet signal at other epochs. The second one uses K-stacker (\citealt{Lecoroller2015, Nowak2018}, Le Coroller et al. 2020, submitted), that is a more generic code for identifying sub-threshold signal in multiple observations of an orbiting object; however, K-stacker will be used here only in a limited way, essentially as a confirmation of the result we obtained with the other approach, with a more extensive use of it being left for the future. The paper is structured as follows: in Section~2 we describe the observations; in Section~3 we outline the methods used in the search and present the results; in Section~4 we discuss our findings in the context of Proxima~c; conclusions and suggestions for further work are in Section 5.

\section{Observations}
 
We list the epochs of the SPHERE GTO observations of Proxima in Table~\ref{tab:obs}. For each epoch, we also give the phases corresponding to the RV orbit of Proxima c, as well as the expected separation along the major axis of the projected orbit. We note here that the orbit proposed by \citet{Damasso2020} is circular. The listed values are the real separation only for an inclination $i=90$~degrees or for a phase of 0.25 or 0.75. For a non-edge-on orbit the actual projected separation will be larger for most orbital phase angles. 

Since Proxima~c is expected to be at a separation larger than 1 arcsec from the star for a large fraction of its orbit, the search should be done over a quite large field of view. We need then to consider data acquired with the IRDIS dual band imager \citep{Dohlen2008}, that has a roughly square field of view with a side of about 11 arcsec, while at best the planet should be in the field of view of the Integral Field Spectrograph (IFS: \citealt{Claudi2008}) for only a fraction of the observations. In order to describe the quality of the observation at each epoch, we give in Table~\ref{tab:obs} the 5-$\sigma$ limiting contrast at 0.5 arcsec obtained with IFS using the deepest analysis method we tried (Principal Component Analysis - PCA - with simultaneous angular and spectral differential imaging: \citealt{2006ApJ...641..556M, Soummer2012, Amara2012, Mesa2015}). All observations were acquired in the IRDIFS mode, that is observing with IRDIS in the H2/H3 dual-band imaging filters \citep{Vigan2010} and with IFS in the Y-J mode.

\begin{table*}
\caption[]{List of SPHERE Proxima observations; contrast is measured at 0.5 arcsec on IFS images; phase c is the phase of planet c from the spectroscopic orbit}
\label{tab:obs}
\centering
\begin{tabular}{lcccccccl}
\hline
MJD & Date & Exp.time & Seeing & Rotation & Contrast & Phase c & Sep (i=90 degree) & Note\\
   &  & s & arcsec & degree &  mag     &         &  mas & \\
\hline 
57112 & 2015-03-31 & ~576 & 0.95 & ~3.2 & 12.18 & 0.6459 &  896 & \\
57406 & 2016-01-19 & 4480 & 2.20 & 25.7 & 13.31 & 0.8007 & 1088 & \\
57436 & 2016-02-18 & 1760 & 1.86 & 13.5 &       & 0.8163 & 1050 & Poor \\
57448 & 2016-03-01 & 3392 & 0.78 & 22.6 & 12.57 & 0.8226 & 1031 & \\
57475 & 2016-03-28 & 4000 & 2.08 & 25.7 & 13.88 & 0.8370 &  984 & \\
57494 & 2016-04-16 & 3840 & 0.62 & 28.7 & 12.65 & 0.8496 &  946 & \\
57830 & 2017-03-19 & 2048 & 1.10 & 11.5 & 13.88 & 0.0237 &  146 & \\
57919 & 2017-06-14 & 2720 & 1.88 & 20.9 & 12.65 & 0.0706 &  467 & \\
58222 & 2018-04-13 & 4320 & 0.58 & 29.3 & 14.23 & 0.2301 & 1127 & \\
58227 & 2018-04-18 & 4480 & 0.55 & 30.0 & 14.00 & 0.2327 & 1130 & \\
58244 & 2018-05-05 & 3840 & 0.65 & 25.7 & 14.31 & 0.2417 & 1137 & \\
58257 & 2018-05-18 & 3840 & 0.40 & 25.7 & 14.36 & 0.2484 & 1139 & \\
58288 & 2018-06-20 & 3840 & 1.70 & 24.1 & 13.88 & 0.2648 & 1137 & \\
58588 & 2019-04-15 & 4608 & 0.47 & 25.8 & 14.55 & 0.4221 &  560 & \\
58621 & 2019-05-17 & 1920 & 0.91 & 26.6 & 14.02 & 0.4401 &  449 & \\
\hline     
\end{tabular}
\end{table*}

   \begin{figure}
   \centering
   \includegraphics[width=\columnwidth]{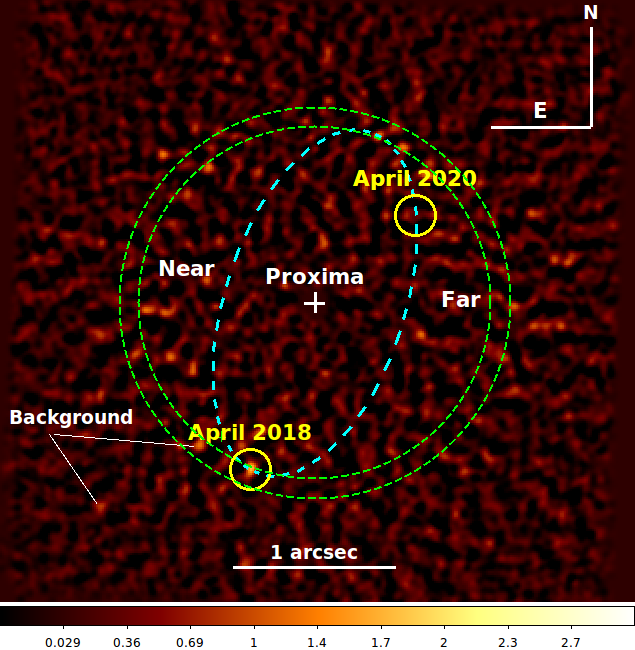}
   \caption{Median of the five epochs of Proxima acquired near quadrature in 2018 (MJD 58222, 58227, 58244, 58257, 58288), combined assuming an inclination of $i=120$~degree. Individual images were rotated to take into account orbital motion with respect to the reference epoch (MJD=58222);  North and East positions are then correct only for that epoch. The search area for c is the ring between the two green dashed circles, with inner radius of 1080 mas and outer radius of 1200 mas. The yellow circles marks the best candidate at the epoch of observation (that is MJD=58222, 2018-04-13) and at mid April 2020; this last is to provide the reader an idea about the speed and direction of the orbit. The cyan dashed line represents the orbit of the candidate planet on the sky plane. Two faint background stars are still visible in this image. The colour bar is the median S/N over the five epochs.}
   \label{fig:comb}
   \end{figure}

\section{Analysis and results}

\subsection{Data preparation}

The reduction of data for individual epochs was performed as described in \citet{Zurlo2018} and makes use of the DRH pipeline \citep{Pavlov2008} as implemented at the SPHERE Data Center \citep{Delorme2017}. We refer the reader to \citet{Zurlo2018} for more details on the pre-processing and the final products of the reduction. We used the images obtained after application of monochromatic PCA to the final images for this analysis (average of those obtained with the H2 and H3 filter of IRDIS), and we removed the background stars inside a radius of 5 arcsec. Background stars with signal above detection threshold ($S/N>5$) in individual images were already identified in the analysis of \citet{Zurlo2018}. The values for pixels where there was signal from the background stars  - there were three such objects in the portion of the images considered in this paper - were replaced by the median of the surrounding background in the data cube prior to PCA analysis.
While removing most of the signal relates to these sources, this procedure may however leave residual ("ghosts") at S/N$\sim 2-3$\ in the corrected images. In addition there are images of two faint background stars (3$<$S/N$<$5, that is, below the threshold for detection in the individual images) still present in the images. They can be separated from noise spikes of similar intensity comparing different images, because they move very rapidly ($\sim 10$~mas/day) over the observed field of view but their relative positions are constant. One of them is out of the considered region in the last epoch. In both cases, "ghosts" of bright and faint background stars may be visible in the median image depending on how their signals combine with local noise at other epochs. Concerning this point, we notice that while at large separation from the star the noise is not strongly correlated pixel-to-pixel, the $S/N$ from a background star is $>2$ over several ($\sim 10$) adjacent pixels because the detector is over-sampled. The probability that a signal at more than $S/N=2$\ appears at the final median image in the position of a background star depends on the noise distribution; we tested this by considering the fraction of pixels that have at least two cases with $S/N>2$ over four extractions out of five, so that the median value combined with the strong residual in the fifth image is $S/N>2$, for at least one pixel over all the pixels with $S/N>2$ corresponding to a background star (that we assumed here to be 10). We found that this probability is $\sim 5$\% if we assume a Gaussian noise distribution; it is larger in a more realistic case where the probability of having peaks at more than 2-$\sigma$ is larger than for a Gaussian distribution. We conclude that having two such cases observed in the median images (as observed, see Figure~\ref{fig:orbit}) is not unlikely.

\subsection{Observations acquired near quadrature}

For a circular orbit with known conjunction epoch, period and semi-major axis, planet position at each epoch depends on two parameters, the inclination $i$\ and the position angle (PA) of the orbit. Combination of the images at generic epochs requires raising the threshold for detection, in order to avoid false alarms. In fact, even small variations of $i$\ and $\Omega$, the longitude of the ascending node, would lead to quite different predictions about the location of the planet at different epochs; combined with the unknown value of orbital PA and the fact that the real orbit is likely not exactly circular, this implies the need to search over a substantial fraction of the available images. However, close to quadrature, that is at phases of 0.25 and 0.75 according to the convention adopted by \citet{Damasso2020}, the difference between the real separation and that along the major axis is negligible for any value of the inclination for a circular orbit. This implies that if we only use observations taken close to this phase, we may limit our search to a narrow ring around the star (see \citealt{Mawet2019} for similar argument). In addition, around quadrature the variation with time of the planet PA along the orbit depends on the inclination at which the orbit is observed, and has an upper limit because the apparent orbital speed should be compatible with the Kepler third law (assuming a mass of 0.12~M$_\odot$\ for Proxima: \citealt{Mann2015}). This leads to a small number of possible combinations (essentially rotations) of the images acquired at different epochs that are compatible with a keplerian motion, and then substantially restricts the volume of the phase space where the search of the Proxima c signal should be done.

Luckily, in 2018 we acquired a sequence of five good data sets, spanning the limited range in phase between 0.2301 and 0.2648, all very close to quadrature. The expected span of separation of planet c image from the star is very narrow, between 1.127 and 1.139 arcsec. This is actually much narrower than the error bar on the semi-major axis from the spectroscopic orbit ($1.137\pm 0.061$~arcsec) and of the Full Width at Half Maximum (FWHM) of the diffraction peak of SPHERE images ($\sim 0.035$~arcsec). We may then assume that during all these epochs, the candidate is at a nearly constant separation, and the planet PA changes with time due to the projection of the circular motion related to the inclination of the orbit. We note that during these epochs, we expect the candidate to be out of the IFS field of view. Hence, we should search for it in the IRDIS images. 

   \begin{figure*}
   \centering
   \includegraphics[width=\textwidth]{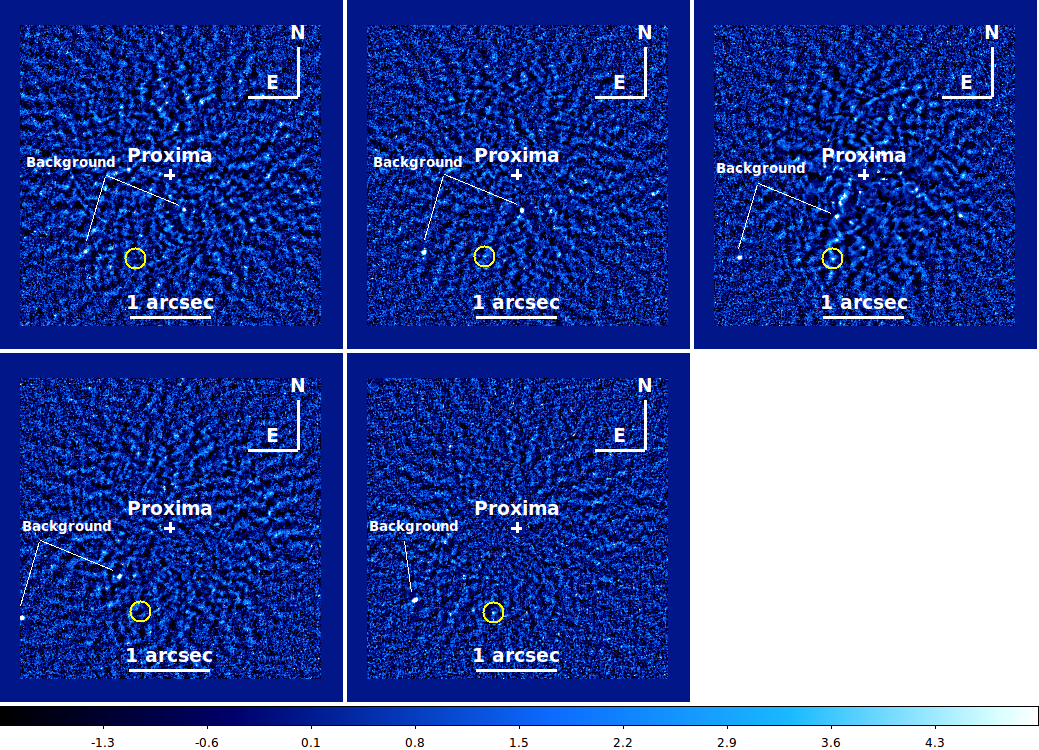}
   \caption{Individual S/N maps for the five 2018 epochs. From left to right: Top row: MJD 58222, 58227, 58244; bottom row:  58257, 58288. The candidate counterpart of Proxima c is circled. Note the presence of some bright background sources not subtracted from the individual images. However, they move rapidly due to the large proper motion of Proxima, so that they are not as clear in the median image of Figure~\ref{fig:comb}. The colour bar is the S/N. S/N detection is at S/N=2.2 (MJD 58222), 3.4 (MJD  58227), 5.9 (MJD 58244), 1.2 (MJD=58257), and 4.1 (MJD58288)}
   \label{fig:ind}
   \end{figure*}

\begin{table}
\caption[]{Measures of the counterpart candidate to Proxima~c at the various epochs}
\label{tab:ind}
\centering
\begin{tabular}{lccc}
\hline
MJD & PA (degree) & Sep (mas) &S/N \\
\hline
57406 & ~~0.08 & 870 & 3.7 \\
57830 & 100.82 & 809 & 1.2 \\
58222 & 157.13 & 1104 & 2.2 \\
58227 & 158.19 & 1068 & 3.4 \\
58244 & 159.36 & 1092 & 5.9 \\
58257 & 160.83 & 1087 & 1.2 \\
58288 & 164.47 & 1079 & 4.1 \\
58588 & 218.67 & 802 & 1.6 \\
58621 & 225.13 & 767 & 3.1 \\
\hline     
\end{tabular}
\end{table}

Following the previous discussion, we combined the signal-to-noise ratio (S/N) maps extracted from the observations acquired in 2018 (spanning about two months) by rotating the images with respect to the first one. This rotation is that expected due to the apparent motion of the planet for different values of the inclination. These S/N maps were obtained with the same procedure presented in \citet{Zurlo2014}, after subtracting obvious background stars. We then made a median of the results, to reduce sensitivity to possible residuals of background objects. We repeated this procedure for different values of the inclination, with a step of 9 degrees in inclination (that is 1.14 degree in field rotation) from 0 to 180 degrees (that is, 21 steps). This step is small enough to ensure that whatever is the correct inclination, we have at least one combination of images for which the difference between the real and model orbital motion leads to a shift of the candidate planet image by less than one pixel (the scale is 12.25 mas/pixel) over the epoch range covered by the 2018 observations. This is about one third of the FWHM of point-source images, which is about three pixels.

In each combined image, we then searched for the highest peak in the ring with separation of $1133\pm 61$~mas, where we expect the companion is close to quadrature; to reduce random variations, we smoothed these combined images using a current median of $3\times 3$~pixels, that is, the expected FWHM of point sources. However, we obtained the same result with different values of the smoothing parameter in the range between 1 (that is, no smoothing at all) to 7 pixels. 

The best candidate that we found with this process is at a separation of 1072 mas (at the inner edge of the search area) and a position angle of PA = 157.90 degrees (for the epoch JD=58222.20). This is obtained for an inclination of $i=120$~degrees (see Figure~\ref{fig:comb}), that corresponds to a counter-clockwise motion, that once combined with RVs implies that the near side of the orbit is on NE and the far one on SW.  We note that the peak corresponding to the candidate found for this angle is the highest in the whole image despite of the fact that the area within the ring covers less than 6\% of the surveyed image. The detection is at a S/N ratio of 6.1 in the median image, estimated as the ratio between the peak S/N and the standard deviation over a $100\times 100$ pixel area at a similar separation from the star. The median value of the S/N values over the images at individual epochs is 3.4. The second highest peak in the median image is at S/N=4.8 and there are only two other peaks above S/N=4, but they are all out of the search area. 

In Figure~\ref{fig:ind} and Table~\ref{tab:ind}, we show the results from the individual epochs. From this data, the S/N ratios for the candidate are quite uniform over the different epochs; though the number of epochs is not large, this is not what expected for a background source, which should not share the large proper motion with Proxima and hence have a high S/N only in a single image. On the other hand, it is clear that only by combining several images we can hope to achieve a reliable counter-identification.

\subsection{How reliable is this detection?}

How much may we be confident that the candidate found using the previous analysis is the counterpart of Proxima c? It is not easy to answer to this question. We considered the issue from different point of views.

\subsubsection{Is the source a random fluctuation?}

First, the probability that this detection is a false positive is possibly low is suggested by the following argument. As mentioned above, the area within the ring is only 6\% of the area of the S/N maps (radius of 2 arcsec), and the candidate peak has the highest S/N in the whole map. In addition, we note that each relative rotation we applied to the images for individual epochs corresponds to create a new noise realization in the combined image so far the rotation between different images is large enough to shift the noise peaks more than their correlation length, that is approximately the FWHM of the PSF. To have a better statistics of the maximum S/N obtained for random peaks, we then extended the rotation over the whole range of 360 degrees, considering also values that are incompatible with keplerian motion (that would limit rotation to the range from -12 to +12 degrees); we did this with a step of 3 degrees to ensure that results obtained for each rotation angle are independent from each other. In this way we searched the entire parameter space, and the sub-threshold solution happens to line up with the parameter space where an orbit is detected. Over all these combinations, we found only one case anywhere in the image where the maximum S/N is higher than for the candidate over the whole combined S/N maps.

There is then 1/120 random instances of signal this high, but the search area is 16 times smaller, so the likelihood of a random signal this high is 1/(120*16),
that is 1 odd over 1920 cases. We may then use a binomial distribution and assume that the number of extractions is equal to the number of independent angles considered for the search of Proxima c (nine) and that the probability of a random result for a single extraction is 2/1920=0.1\% (counting also the possible detection of Proxima c candidate because we should consider the possibility that this detection is a random result). The result is a probability of about 0.9\% of finding by chance a peak as strong as that observed within the search area (distance from the star and field rotation) considered. Statistics dealing with one detection are always tricky: for instance, this test assumes that the distribution of noise in the S/N maps is the same over the whole field of view, a fact that is likely not true close to the star (see e.g. \citealt{Mawet2014} and references therein), and that noise realizations are really independent from each other, possibly not true close to the star. It is then possible that the false alarm probability is higher than estimated by this test. Anyhow, this low probability suggests that we have found an interesting candidate for the optical counterpart of Proxima c.

\subsubsection{Is the source in a region of high background noise?}

Another point of possible concern is the following. In the image obtained at JD 58244 there appears to be a feature starting to the left of Proxima Cen and passing through the position where the Proxima c candidate signal is claimed. This is the highest S/N detection of the companion, and the possibility of the detection being a peak in this linear feature (presumably a correlated noise feature) should be considered. We then had a closer look to this structure. It includes both a background object and the possible image of Proxima c candidate. To avoid biasing the results due to these features, we only considered the region between them and avoiding regions closer than 5 pixels to either of them. This comparison region includes about 600 pixels. The average S/N value within it is $0.10\pm 0.05$, that is, slightly different from zero (that is the expected value) but only at about 2$\sigma$\ level. The rms within this region is $1.25\pm 0.05$ (in noise units). The region is then slightly more noisy than the whole image. The significance of these two results is not clear, because this area was picked out among the images obtained at different epochs exactly because it looked a bit different at eye inspection. However, the differences in both the mean value and rms scatter are much lower than the S/N measured for the candidate position (that is, 5.9). On the other hand, this result suggests that the S/N obtained for the possible position of Proxima c candidate in this image might be overestimated by some 25\%, because the local noise is possibly underestimated. This may contribute to explaining why this value deviates so much from the typical value of about 3 obtained at other epochs. In any case, the impact of this result in our conclusion is marginal, because we are using median values over the different images. If we neglect this image, the median value reduces from 3.4 to 2.8.

\subsubsection{Is the S/N fluctuation as expected?}

A more general concern is whether the observed fluctuations of the S/N among different observations are compatible with the hypothesis of a real signal. 

In general, we expect that:
\begin{itemize}

\item  there is a scatter in the S/N detections over different visits, even if the signal and noise were constant. This is due to the actual realization of noise in the individual observations. If the S/N is correctly evaluated, not an easy task in high contrast imaging (see also the discussion above), we expect that if the noise is only due to the background, the rms of the S/N in single observation should be 1, and its measure should actually be $1.0\pm 0.3$\ because we are considering 9 epochs. On the other hand, if the noise is mainly due to the source itself, the rms of the S/N should be roughly equal to $\sqrt(<S/N>)=1.7\pm 0.3$. If we consider all the values listed in Table~\ref{tab:ind}, the mean value for the S/N is 2.9 with an rms equal to $1.5\pm 0.5$ (the rms reduces to $1.3\pm 0.4$\ if the S/N for epoch 58257 is reduced to 4.6, as discussed above). This value is intermediate between that expected for background and source noise dominated cases, and compatible with both. This looks appropriate for this region of the image, further from the star than the high noise ring related to the outer working angle of the adaptive optics. Given the uncertainties in these determinations, this result is reasonable, though of course we cannot exclude the possibility that noise is slightly underestimated (or simply that signal and/or noise are not constant). 

\item We might expect that there should be some correlation of the S/N of detection with the (nominal) limiting magnitude (under the hypothesis that the signal is constant). There is actually only a very weak negative correlation (Persson correlation coefficient r=-0.155). Is this unexpected? First, we should notice that the 2016 data sets, that are of much lower quality - are combined together in Table 2. If we combine quadratically the limiting value for each epoch (taking into account they are logarithmic quantities), the appropriate limiting magnitude for the combined 2016 observation should be read as 14.12 mag, very similar to the average for the other epochs listed in Table~\ref{tab:ind} (14.16 mag). Second, the values of the nominal limiting magnitudes are also affected by uncertainties and do not change greatly within the sample. The rms of the values is 0.23 mag, a value that is actually close to the uncertainties in the photometric zero points in observations with SPHERE (0.18 mag) - this point is discussed at some length in Langlois et al. (2020, in preparation), where we discuss the accuracy of the photometry of the SHINE survey. This uncertainty is related to the variations of the Strehl ratio between the observations obtained with the star out of the coronagraphic mask - used to determine the zero point of the photometry - and the science exposure with the star behind the coronagraph. Note that this is likely an underestimate of uncertainties in the limiting contrasts because it neglects any other effect (such as errors in the estimate of correct noise level). 

If we combine uncertainties in the limiting magnitude and in the S/N of the source, we really do not expect that the correlation of the S/N of detection with the (nominal) limiting magnitude should be obvious in the data.  
To show this, 
we considered the case where there is actually a real signal as given by the mean value of the observations (contrast of 14.75 mag), and that the observations have a real limiting magnitude with a Gaussian distribution with a mean value of 14.15 mag and an rms scatter of 0.15 mag over a set of 9 observations. 
We then added realistic Gaussian noise both in the S/N estimate ($\sigma=1.5$) and on the limiting magnitude estimate ($sigma=0.18$~mag), reproducing the observed scatter in S/N values and limiting magnitude (0.23 mag). We then repeated this procedure 10000 times and looked for how many cases the Pearson correlation coefficient between limiting magnitude and S/N is below the observed value  r=-0.155. We found that 
if there is a real signal but in the presence of realistic noise estimates, in 20\% of the cases we should have found a negative correlation stronger than we observed between limiting magnitude and S/N.
\end{itemize}i) 

Summarizing, these arguments do not show that we have detected Proxima c, but simply that the S/N fluctuations among different observations are compatible with this hypothesis - even assuming that the signal is constant.

\subsection{Results obtained with K-stacker}

K-Stacker (\citealt{Lecoroller2015, Nowak2018}, Le Coroller et al. 2020, submitted) is a method of observation and reduction that consists in combining high contrast images recorded during different nights, accounting for the orbital motion of the putative planet that we are looking for, and then looking for peaks in the final S/N maps obtained in this way. The K-Stacker approach takes long computation time because this brute-force algorithm searches simultaneously for new planets and for their orbital parameters (Le Coroller et al. 2020, submitted). A full K-Stacker solution using all epochs is not included in this paper; we plan to present it in a forthcoming paper. Le Coroller et al. (2020, submitted) show how it is important that a stacking approach is used on data of homogeneous (high) quality; for this reason, for this first approach we used only the 2018 epochs and we limited the exploration to those range of orbital solutions that are consistent with the spectroscopic orbit ($1.2<a<1.5$~au and $e<0.1$). We assumed a stellar mass of $m=0.135$~M$_\odot$, that is the mean of the values given by \citet{Mann2015} and \cite{Zurlo2018}. The mass used in the K-stacker computation is not the best value, but K-stacker computations are very time demanding. Since this assumption  does not not play a central role in the paper, we prefer to avoid repeating these computations.

K-Stacker always gives a family of solutions (Le Coroller et al. 2020, submitted). The solution we obtained with the highest S/N is the same peak found in the quadrature analysis described in the previous subsection and also the orbital parameters are very similar ($a=1.47$~au, $i=122$~degree, $\Omega=165$~degree, and a moderate eccentricity of $e=0.066$). The value of the S/N=4.1 is lower than in the quadrature analysis but since it is computed using a different algorithm, there is no reason for the two values to coincide exactly. More importantly, the S/N value is below the threshold usually considered for reliable orbits (see \citealt{Nowak2018} and Le Coroller et al. 2020, submitted). We have however to consider that, compared with the cases considered in those papers, in the present case we are exploring only a limited region of the possible parameter space where keplerian solutions may fall, reducing the probability of false alarms. Overall, while this result is not a really independent confirmation of the detection of c on SPHERE data, it at least confirms that the family of solutions considered in the previous subsection is the one that best matches the 2018 SPHERE observations in the framework of orbits satisfying the constraints set by the spectroscopic orbit.

\subsection{Other epochs}

\begin{figure}[htb]
\centering
\includegraphics[width=\columnwidth]{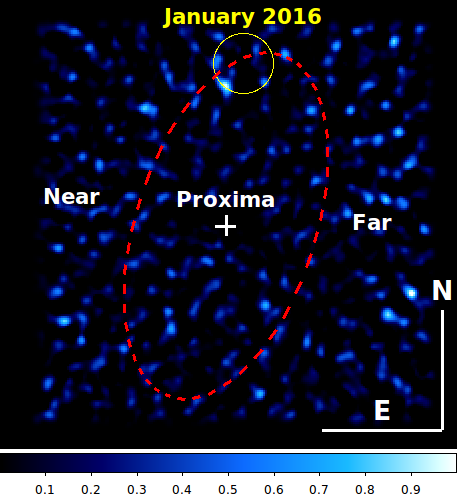}
   \caption{Median of five 2016 observations, shifted to the first of these epochs (January 2016: JD=57406) according to the circular orbit suggested by 2018 observations (red ellipse). The yellow circle is the expected position at that epoch; the radius of this circle is $\sim 180$~mas.  The colour bar is the median S/N over the five epochs.}
\label{fig:2016}
\end{figure}

\begin{figure}[htb]
\centering
\includegraphics[width=\columnwidth]{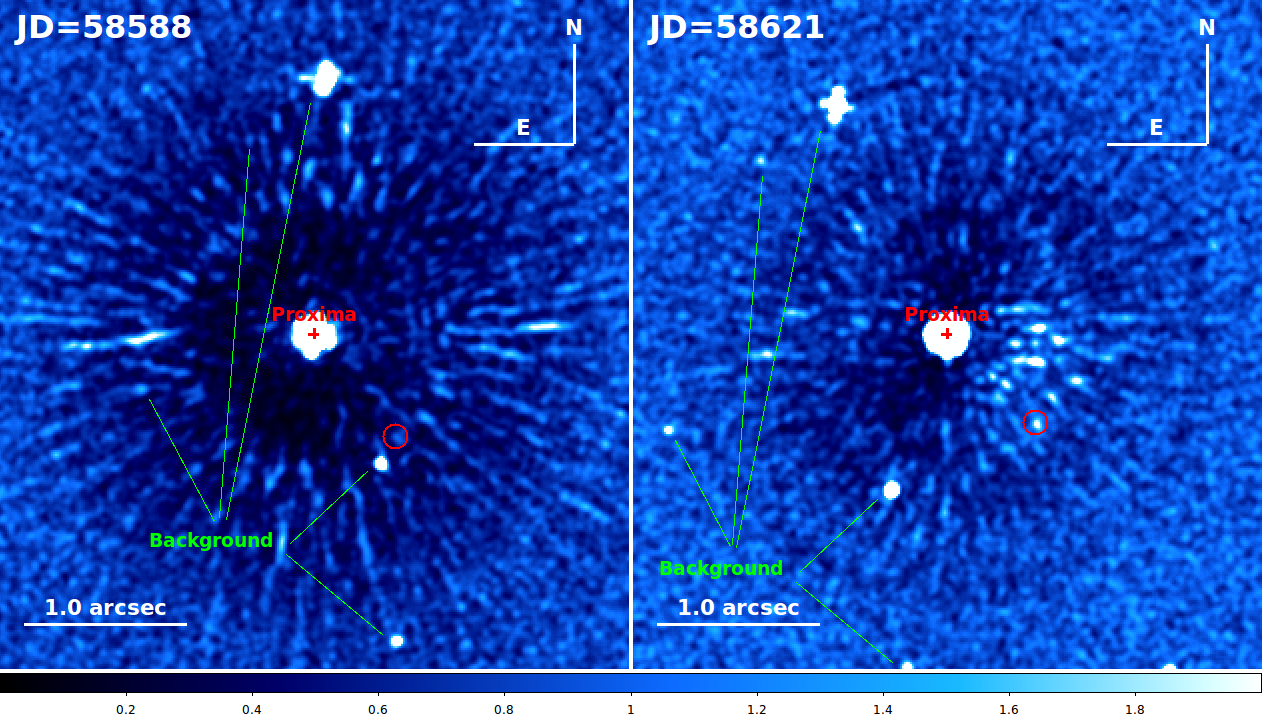}
   \caption{S/N maps for the two epochs acquired in 2019 (JD=58588 and 58621). The red circles mark the possible counterparts of Proxima c at these epochs. Background stars are also shown. The colour bar is the S/N.}
\label{fig:2019}
\end{figure}

\begin{table}[htb]
\caption{Proposed orbital parameters}
    \centering
    \begin{tabular}{lccc}
    \hline
    Parameter & Spectroscopic & Circular & Elliptical \\
              &    orbit      &   orbit  &   orbit    \\
    \hline          
Period (yr)                & $5.20\pm 0.26$ &   5.20  & $5.08\pm 0.34$   \\
Epoch of quadrature (JD)  & $58260\pm 100$ &  58260  & $58165\pm 80$    \\
Eccentricity              &      0.0       &   0.0   & $0.080\pm 0.044$ \\
Semi-major axis (arcsec)  & $1.14\pm 0.06$ & $1.072$ & $1.02\pm 0.06$   \\
Semi-major axis (au)      & $1.48\pm 0.08$ & $1.40$  & $1.33\pm 0.08$   \\
$\Omega$ (degrees)        &                &   164   & $150\pm 7$       \\
$\omega$ (degrees)        &                &         & $-2\pm 34$       \\
Inclination (degrees)     &                &   120   & $137.3\pm 6.2$   \\
    \hline 
    \end{tabular}
    \label{tab:param}
\end{table}

Only two observations of rather poor quality were acquired in 2017 when the planet was likely at a smaller separation from the star and possibly close to the noisy region around the outer working angle of the adaptive optics. On the other hand, we may look for the candidate derived from 2018 data in the sequence of observations acquired in 2016. This data set includes several observations but they were not acquired close to quadrature and are of lower quality and due to worse sky conditions. We retrieved the five best of these images – those from January to April 16 (MJD=57406, 57436, 57448, 57475, 57494). We then combined these images by shifting them to the first epoch as expected from the circular orbit that best matches the 2018 observations and then making a median (this procedure is similar to that described by \citealt{Showalter2019} in their search for the seventh moon of Neptune). The result is shown in Figure~\ref{fig:2016}. There is not a maximum at the expected position coordinate. The closest one (S/N=3.7, the second highest peak in the whole S/N map) is at about 15 pixels ($\sim 180$~mas) from the expected position. This might be due to three reasons: (i) The candidate we retrieved in 2018 is not the counterpart of Proxima c; (ii) Proxima c was fainter in 2016; this is not unlikely because we expect that at this epoch visibility was less favorable, the planet possibly being on the near side of the orbit and then showing a larger fraction of its dark side; (iii) the orbit we assumed is not the correct one. For instance, if we only had adopted a period at the longest edge of the error bar (leaving aside other uncertainties as e.g. a small but not zero eccentricity), we would have expected Proxima c to be at about 100~mas from the expected position for the nominal period. On the other hand, we might assume that this local maximum is a real detection of Proxima c (yielding a position of dRA=1~mas and d$\delta$=870~mas, at JD=57406).

We also explored the two epochs acquired in 2019 (JD=58588 and 58621) that are of fairly good quality (Figure~\ref{fig:2019}). Since these epochs were acquired far from quadrature, the expected location of Proxima c strongly depends on the orbit inclination. The most reasonable counter-identifications are at sep=802 and 767 mas and PA=218.7 and 225.2 degrees, with S/N=1.6 and 3.1, for JD=58588 and 58621, respectively. This is further from the star than expected for an inclination of 120 degrees, and suggests a lower inclination of the orbit. However, there are other peaks in the images of comparable strength, though incompatible with the orbit of Proxima c. Again, this underlines that we are unable to have unambiguous identification of the counterpart of Proxima c on any single epoch.

\begin{figure*}[htb]
\centering
\includegraphics[width=\textwidth]{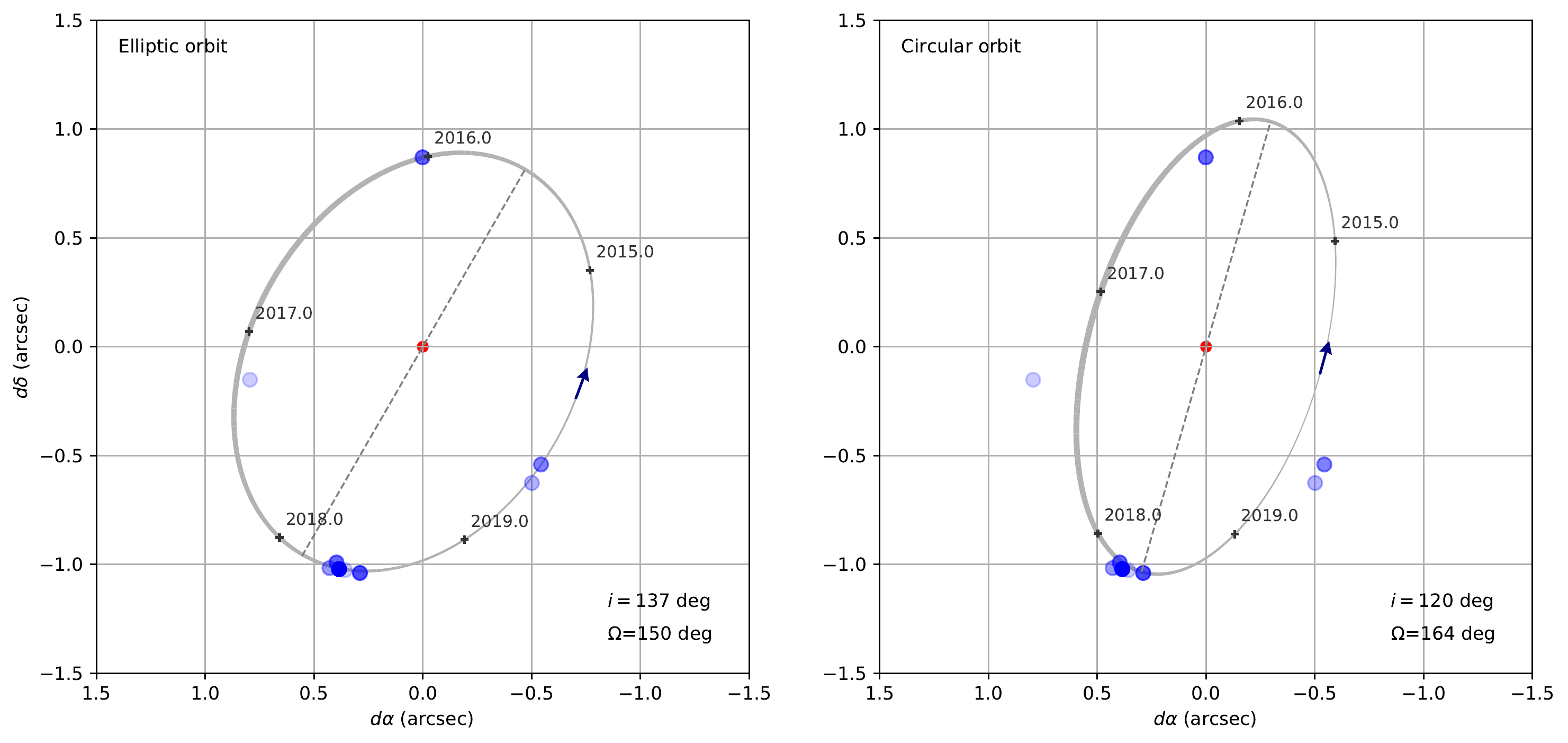}
   \caption{Comparison between the proposed orbits for the candidate counterpart of Proxima~c and the observations listed in Table~\ref{tab:ind}. The density of the blue color of the points is proportional to the S/N in the individual epochs. Left panel: eccentric orbit; Right panel: circular orbit}
\label{fig:orbit}
\end{figure*}

Using all epochs but 2017 (that is very uncertain; see Table~\ref{tab:ind}), we may look for an eccentric orbit solution. We assumed that position errors are equal to 0.04/S/N arcsec \citep{Zurlo2014} and run the ORBIT fitting code by \citet{Tokovinin2016}\footnote{https://zenodo.org/record/61119\#.Xg83GxvSJ24} that is based on a Levenberg-Marquard optimization algorithm to find the best astrometric orbit. The best solution (see Figure~\ref{fig:orbit}) has a period of $P=5.08\pm 0.34$~yr, T0=$2015.56\pm 0.22$, eccentricity $e=0.080\pm 0.044$, semi-major axis $a=1.02\pm 0.06$~arcsec, the angle of descending node $\Omega=150\pm 7$~degrees, the  argument of periastron $\omega=-2\pm 34$~degrees, and $i=137.3\pm 6.2$~degrees, with a reduced $\chi^2=1.35$. This orbit agrees well with that from RVs see Table~\ref{tab:param}).\footnote{The mass of Proxima given by this orbit is $M=0.091\pm 0.012$~M$_\sun$, to be compared with the value of $0.12\pm 0.02$~M$_\sun$ given by \citealt{Mann2015}, that was adopted when determining the best circular orbit from 2018 data alone (Section 3.2). The lower value for the mass corresponding to the eccentric orbit is correlated to its lower inclination. However, the difference with the value given by Mann et al. is within the respective error bars.}


We note that while the orbital inclinations obtained using the circular and eccentric approach appear inconsistent with each other, this is not a real difference. In fact, the circular solution assumes the mass of the star, information from RVs (the range of semi-major axis and the phase), and only the astrometric data for 2018. What was important here was to reproduce the apparent speed of Proxima c candidate projected on sky during this epoch. This speed can be obtained using a high inclination, but it can also be reproduced using a lower inclination but a lower stellar mass and/or assuming an eccentric orbit, with the 2018 observations being acquired not far from apoastron. There is a degeneracy in this astrometric solution that can be broken using a wider set of measurements (2016, 2018, and 2019), that incidentally provides a dynamical estimate of the mass of Proxima and does not use RVs. The assumption of a circular orbit is however crucial in our analysis both to have a first guess for the position angle on sky and to estimate the probability that the detection of Proxima c candidate is a false alarm due to a random alignment of noise peaks.

While these are all reasonable explanations, we still think that the result is ambiguous and stress the need for further observations (direct imaging, radial velocities, and astrometry) to set the counter-identification of Proxima c on more solid basis.

\section{Discussion}

\subsection{Mass and luminosity of Proxima c candidate}

With an inclination of $i=120$~degrees (circular orbit), the mass of Proxima c candidate from the spectroscopic orbit would be $7.2\pm 2.2$~$M_\oplus$; if the inclination is $i=137.3$~degrees (eccentric orbit), the mass is $8.6\pm 2.8$~$M_\oplus$. 

The average contrast measured for the candidate of Proxima c from SPHERE images is $(3.5\pm 2.0)\times 10^{-7}$, that is about three times fainter than the upper limit found by \citet{Mesa2017}. Intrinsic emission is surely negligible at the age of Proxima (4.8 Gyr: \citealt{Thevenin2002, Bazot2016, Kervella2017}); the planet should in fact have a mass of about 5~M$_{\rm Jupiter}$ to be so bright, using AMES-COND models \citep{Allard2011}. We can then focus on the reflected star light. In this case, the contrast is:
\begin{equation}
c = \phi A~r^2/(4~d^2),
\end{equation}
where $A$ is the albedo, $r$ is the radius of the planet, $d$ the distance from the star (1.3 au=$2\times 10^8$~km), and $0<\phi<1$ a parameter that takes into account the fractional illumination and scattering function along the orbit and depends on orbital inclination and phase. Hence, the observed contrast implies a radius of:
\begin{equation}
r = 2~d~\sqrt{c/(\phi A)} = 4E8 \sqrt{(3.5\pm 2.0)\times 10^{-7} /(\phi A)}.
\end{equation}
If $\phi A\sim 0.5$, we obtain $r\sim (3.4\pm 1.2)\times 10^5$~km, that is about $4.8\pm 1.7$~R$_{\rm Jupiter}$. This is at least an order of magnitude too large for a Neptunian planet with a mass of 5-11~$M_\oplus$. However such a large size is possible for a system of rings or dust clouds around the planet; a similar explanation has been considered for Fomalhaut b \citep{Kalas2008, Kennedy2011, Janson2012, Tamayo2014}; see however \citet{Galicher2013, Lawler2015} for a different interpretation of Fomalhaut b, and \citet{Kalas2018} for a more recent discussion. 

In this context, so long as our detection of Proxima c is not spurious, we note that it appears unresolved in SPHERE images, with a FWHM of $\sim 35$~mas, yielding an upper limit of about 120~R$_{\rm Jupiter}$ to the emitting area of the dust cloud. This value is consistent with the Hill radius (expected to be $39\pm 2$~mas) and with the size of satellite systems of giant planets in the Solar System. Furthermore, the dynamical mass for Proxima c candidate provided by RVs makes it consistent with the minimum mass (a few M$_\oplus$) required for exciting the collisional cascade in the swarming satellites scenario considered by \citet{Kennedy2011}. Similar information is not available for Fomalhaut b. 

On the other hand, the age of Proxima (4.8 Gyr: \citealt{Thevenin2002, Bazot2016, Kervella2017}) is similar to that of the Sun, and much older than Fomalhaut ($440\pm 40$~Myr: \citealt{Mamajek2012}). For what we know at present, Proxima possibly contains only one long period planet, to be compared with the four giant planets of the Solar System, so that a planetary system around Proxima might be dynamically less evolved than the Solar System. However, at present we are not able to be sure about this. When compared to the Saturn ring system, the required disk system has a radius of $\sim 5$~R$_{\rm Jupiter}$ or larger, while the bright portion of the Saturn rings (rings B and A) extends up to about 2~R$_{\rm Jupiter}$, and the tenuous E-ring extends up to almost 7~R$_{\rm Jupiter}$. So the disk system required to explain the observation of Proxima c candidate should be likely larger than that of Saturn, but not by orders of magnitude. Why there are disks or ring systems around old planets is not known, and hence it is difficult to estimate the likelihood of observing them around a putative Proxima c. In the case of Saturn, mainly two scenarios are considered in the literature. In the first one the rings are residuals of an original much more massive disk that formed early in the system, either at formation or during the Late Heavy Bombardment \citep{Canup2010, Charnoz2011}. In the second one, largely based on the results of the Cassini mission, the rings are temporary features that can live a few hundred Myrs and they are then relatively young \citep{Ida2019, Iess2019}. Still they are created after collisions and disruptions of massive bodies, that would be more probable in the early phases of the Solar System evolution \citep{Kerr2008}, but it is not clearly impossible at older ages. As a reference of the size of the bodies involved, the mass of Saturn's rings is estimated to be roughly half the mass of Mimas \citep{Iess2019}, that has a radius of about 200 km.

\subsection{Comparison with \textit{Gaia} results}

If Proxima~c exists, it should be detectable with the next \textit{Gaia} data releases (see \citealt{Damasso2020}). Given the planet (5-11~$M_\oplus$) and host ($0.12\pm 0.02$~M$_\sun$, \citealt{Mann2015}) masses and the semi-major axis of the orbit ($1.01\pm 0.05$~arcsec), the astrometric signal should be $\sim 200~\mu$as, with a period of 5.2~yr. This is well above the detection limit of \textit{Gaia} \footnote{See https://www.cosmos.esa.int/web/Gaia/science-performance }.

In the meantime, we may compare our proposed orbit with the astrometric signal that is obtained using \textit{Gaia} DR2. The evidence for companions to Proxima on this data has been explored by \citet{Kervella2019} and by \citet{Brandt2018, Brandt2019}
\footnote{The saturation with Gaia starts around $G=12$ \citep{Evans2018}), while Proxima has $G=9$. The star is relatively faint in the visible, but its flux in the red is higher and the Gaia filter is very broad. This level of flux ($G=9$) does not pose particular problems with the Gaia astrometry. This is shown by the RUWE (Renormalized Unit Weight Error) of Proxima that is equal to 1.0 \citep{Kervella2019} which is in the reliable range ($<1.4$). This shows that the star is behaving correctly, compared to other stars of similar brightness.}, who obtained consistent results.
In the following, we will consider more specifically the results obtained by \citet{Kervella2020} (similar result are obtained by \citealt{Benedict2020}). By comparing the short-term proper motion vector measured by \textit{Gaia} DR2 with the long term trend derived using the \textit{Gaia} DR2 and Hipparcos positions, \citet{Kervella2019} concluded that during the epoch surveyed by \textit{Gaia} DR2 (2014 to 2016) Proxima exhibited a proper motion anomaly (hereafter PMa, compared to the long-term proper motion) of $d\mu_{\rm RA}= +0.22\pm 0.11$\ and $d\mu_{\rm Dec}= +0.38\pm 0.21$~mas/yr, that is, toward North East.
Since the mass ratio between Proxima and planet~b is likely $>10^4$ and the maximum separation $<40$~mas, the reflex motion of Proxima around the barycenter has an amplitude of only $<4$~$\mu$arcsec. This is by far not enough to significantly affect the \textit{Gaia} astrometry, so the observed PMa signal is entirely caused by planet~c or additional massive long-period orbiting bodies.

While the error bars are still large, so that the significance of the PMa is only at a 1.8~$\sigma$ level, we may compare it with the prediction for the orbit we determined for the planet. We find that planet~c moved toward NE with respect to Proxima during the \textit{Gaia} DR2 time window (around J2015.5), at an average rate of $d\mu_{\rm RA}=+0.66$\ and $d\mu_{\rm Dec}=+0.54$~arcsec/yr. Given the mass ratio between the planet and the star, this yields a proper motion anomaly of $d\mu_{\rm RA}=-0.20$\ and $d\mu_{\rm Dec}=-0.15$~mas/yr for Proxima. This is opposite to what is found from the \textit{Gaia} DR2 by \citet{Kervella2019,Kervella2020}. As a possible explanation of this difference of $\approx 3\sigma$, another massive planet may be orbiting Proxima with a longer period. Additional Gaia astrometric measurements are required to obtain a clear detection. The \textit{Gaia} (E)DR3 will contain only time averaged PM values and not individual epochs (as the DR2). However, it should be possible from the combination of DR3 and DR2 to detect the signature of Proxima c (and putative additional bodies) at a significance level of more than $3\sigma$.


\subsection{Comparison with ALMA results}

The inclination ($|i|=-60$ degrees) and position angle of the circular orbit given by this candidate optical counterpart (162 degrees) are not too far (misalignment $<30$ degrees) from those proposed for the outer belt at $\sim$30 au (inclination of 45 degree and position angle of 140 degree) and for the cold ring (deconvolved size of $\sim 0.8$~au, PA$\sim$130 degrees) proposed by \citet{Anglada2017} from ALMA 12m array data. The agreement is even better (misalignment of 14 degrees, with a probability of chance alignment of $\sim 3$\%) if we consider the tentative eccentric orbital plane ($\Omega=150\pm 7$~degrees, and $|i|=42.7\pm 6.2$~degrees). This is remarkable in view of the uncertainties in both estimates. However, it is quite possible that the orbit of c is not exactly coplanar with the outer belt and cold ring.

\citet{Anglada2017} also proposed a secondary source from the ALMA 12m array data at a separation of 1.2 arcsec and PA=114 degree, at MJD=57869. For comparison, our candidate counter-identification of Proxima c is at 1.08 arcsec and PA=158.4 degrees at MJD=58222. The question naturally rises whether they could be the same object. Assuming our proposed circular (values in parenthesis are for the eccentric) orbit, we expect that the Proxima c candidate to be at sep=0.59 (0.91) arcsec and PA=100 (108) degrees at the epoch of the observation by \citet{Anglada2017}. Given that the ALMA beam during this observation was 0.7 arcsec and that we have an uncertainty of 6 degrees in the PA of the circular orbit, the PA difference of 14 degree (6 degree for the eccentric orbit) is well within the uncertainties. The probability that this alignment is obtained by chance is 3.9\% for the result of the circular orbit, and 1.7\% for that of the eccentric orbit. On the other hand, the expected separation of Proxima c candidate at the epoch of \citet{Anglada2017} observation is lower than that found on the ALMA data for both circular and eccentric orbits (in this second case, by only $\sim 0.3$~arcsec). \citet{Anglada2017} discuss the possibility that the emission observed by ALMA 12m array results from a ring containing some $10^{-5}$~M$_\oplus$ of dust around a planet, and cite theoretical arguments that the planet should be $10^7$ times more massive \citep{Charnoz2018}; this is about ten times larger than the putative case of Proxima c. Given the large size and mass, it is possible that these counterparts to the RV signal are due to a ring or a debris disk, or it could be also a post-collisional disk (planet-planet collision). Alternatively, dust around Proxima c could be generated by collisional evolution of satellite swarms \citep{Kennedy2011}.

It is also possible that the millimeter emission detected by \citet{Anglada2017} could be a cometary tail consisting of small size particles driven by the stellar wind. This might not be unexpected given the high level of activity of Proxima (see e.g. \citealt{Garraffo2016, Ribas2017, MacGregor2018, Howard2018}) and the possible presence of dust around the companion. 

Finally, it is also possible that the emission observed by \citet{Anglada2017} has a very different explanation. 
%
%
For instance, it may be a background source, such as a distant galaxy, although the probability of observing such a source at 1.2 arcsec from Proxima is $\leq 10^{-2}$\ \citep{Fujimoto2016ApJS, Anglada2017}. Also, current data cannot completely rule out the possibility it is just a noise peak.
Further observations are clearly needed to clarify this subject.

\section{Conclusions}

While we are not able to provide a firm detection of Proxima c, we found a possible candidate that has a rather low probability of being a false alarm. If our direct NIR/optical detection of Proxima c is confirmed (and the comparison with early \textit{Gaia} results indicates that we should take it with extreme caution), it would be the first optical counterpart of a planet discovered from radial velocities. A dedicated survey to look for RV planets with SPHERE lead to non-detections \citep{2018MNRAS.480...35Z}. If real, the detected object (contrast of about 16-17 mag in the H-band) is clearly too bright to be the RV planet seen due to its intrinsic emission; it should then be circumplanetary material shining through reflected star-light. In this case we envision either a conspicuous ring system \citep{Arnold2004}, or dust production by collisions within a swarm of satellites \citep{Kennedy2011, Tamayo2014}, or evaporation of dust boosting the planet luminosity (see e.g. \citealt{Wang2019}). This would be unusual for extrasolar planets, with Fomalhaut b \citep{Kalas2008}, for which there is no dynamical mass determination, as the only other possible example. Proxima c candidate is then ideal for follow-up with RVs observations, near IR imaging, polarimetry, and millimetric observations. Due to the strong interest among the community and the public (see Proxima b, and the Breakthrough Starshot program: \citealt{Kipping2017} and its erratum \citealt{Kipping2018}), the confirmation would be an important achievement for the field.

In addition, we note that if our detection is true, and the orbit of b is coplanar with that of c (consistent with the small misalignment between the best orbit for c and the outer belt revealed by ALMA: \citealt{Anglada2017}), the mass of Proxima b would be about 1.5-1.8~M$_\oplus$. An inclination of $|i|=42-60$~degrees should not lead to transits of b (in agreement with observations of \citealt{Feliz2019}) for any reasonable value of the planet radius. 

Given the potential relevance of this work and its possible influence on the future detection of b, it should be urgently confirmed by new high contrast observations. 

\begin{acknowledgements}
 R.G., D.M., S.D., and M.D. acknowledges financial support from Progetto Premiale 2015 FRONTIERA (OB.FU. 1.05.06.11) funding scheme of the Italian Ministry of Education, University, and Research. A.Z. acknowledges support from the CONICYT + PAI/ Convocatoria nacional subvenci\'on a la instalaci\'on en la academia, convocatoria 2017 + Folio PAI77170087. E.R. and R.L. are supported by the European Union’s Horizon 2020 research and innovation programme under the Marie Skłodowska-Curie grant agreement No 664931. J.H. is supported by the Swiss National Foundation (SNSF) for this work through the $\#PZ00P2_180098$ grant. This work has been supported by the project PRIN INAF 2016 The Cradle of Life - GENESIS-SKA (General Conditions in Early Planetary Systems for the rise of life with SKA). The authors acknowledge financial support from the Programme National de Plan\'etologie (PNP) and the Programme National de Physique Stellaire (PNPS) of CNRS-INSU. This work has also been supported by a grant from the French Labex OSUG\@2020 (Investissements d’avenir - ANR10 LABX56). The project is supported by CNRS, by the Agence Nationale de la Recherche (ANR-14-CE33-0018). This work is partly based on data products produced at the SPHERE Data Centre hosted at OSUG/IPAG, Grenoble. We thank P. Delorme and E. Lagadec (SPHERE Data Centre) for their efficient help during the data reduction process. SPHERE is an instrument designed and built by a consortium consisting of IPAG (Grenoble, France), MPIA (Heidelberg, Germany), LAM (Marseille, France), LESIA (Paris, France), Laboratoire Lagrange (Nice, France), INAF Osservatorio Astronomico di Padova (Italy), Observatoire de Genève (Switzerland), ETH Zurich (Switzerland), NOVA (Netherlands), ONERA (France) and ASTRON (Netherlands) in collaboration with ESO. SPHERE was funded by ESO, with additional contributions from CNRS (France), MPIA (Germany), INAF (Italy), FINES (Switzerland) and NOVA (Netherlands). SPHERE also received funding from the European Commission Sixth and Seventh Framework Programmes as part of the Optical Infrared Coordination Network for Astronomy (OPTICON) under grant number RII3-Ct-2004-001566 for FP6 (2004-2008), grant number 226604 for FP7 (2009-2012), and grant number 312430 for FP7 (2013-2016). This work is supported by the French National Research Agency in the framework of the Investissements d’Avenir program (ANR-15-IDEX-02), through the funding of the "Origin of Life" project of the Univ. Grenoble-Alpes. 
\end{acknowledgements}

\bibliographystyle{aa} 
\bibliography{main.bib} %

\end{document}